\documentclass[amsmath,amssymb,twocolumn]{revtex4-2}
\usepackage{CJK} 
\usepackage{bm}

\usepackage{amsmath,amssymb} 
\usepackage[usenames,dvipsnames]{color}
\usepackage{multirow}

\usepackage{graphicx} 
\usepackage{bm}
\usepackage{relsize}
\usepackage{braket}

\usepackage{float}
\usepackage[unicode=true,colorlinks=true,urlcolor=blue,citecolor=blue]{hyperref}
\usepackage{hyperref}
\usepackage[normalem]{ulem}

\begin{document}
\begin{CJK*}{UTF8}{gbsn} 
\title{Symmetry Classification for Alternating Excitons in Two-Dimensional Altermagnets}

\author{Jiayu David Cao (曹嘉宇)}
\thanks{Present address: University of Central Florida,
Orlando, Florida, USA.}

\author{Konstantin S.~Denisov}

\author{Yuntian Liu (刘云天)}
\email{yliu369@buffalo.edu}

\author{Igor \v{Z}uti\'{c}}

\affiliation{Department of Physics, University at Buffalo, State University of New York, Buffalo, New York 14260, USA} 
\begin{abstract}
Excitons, bound electron-holes states, often dominate the optical response of two-dimensional (2D) materials and reflect their inherent properties, including spin-orbit coupling, magnetic ordering, or band topology. By focusing on a growing class of collinear antiferromagnets with a nonrelativistic spin splitting, referred to also as altermagnets (AM), we propose a theoretical framework based on the spin space group (SSG) to elucidate their resulting excitons. Our approach is illustrated on 2DAM with spin-polarized valleys, where we classify the combination of conduction and valence bands by the SSG representations into two cases that hosts bright $s$-like and $p$-like excitons, respectively. This analysis is further supported by effective Hamiltonians and the Bethe-Salpeter equation. We identify the 
excitonic
optical selection rules from the calculated absorption spectra and the symmetry of bright excitons from their momentum space envelope functions. Together with first-principles calculations, several material candidates are predicted for realizing excitons in 2DAM.
Our framework provides optical fingerprints for various cases of AM, while their tunability, such as the strain-induced valley splitting, is also transferred to excitons allowing, additionally, valley-polarized photocurrent generation.  
\end{abstract}
\maketitle
\end{CJK*}

A growing class of two-dimensional (2D) materials is characterized by reduced dielectric screening which enhances Coulomb interactions and many-body 
effects~\cite{Zhang2005:N,Burch2018:N,Huang2022:NN,Grzeszczyk2023:AM,Wang2015:NN,VanTaun2017:PRX,Tran2014:PRB,Scharf2019:PRB,Shao2025:NM}.
The resulting excitons, as exemplified in 2D transition metal dichalcogenides (TMDs)~\cite{Wang2018:RMP}, can propagate over macroscopic 
distance~\cite{Unuchek2018:N,Zhou2024:NC}, have orders of magnitude larger binding energy than their bulk counterparts~\cite{Yu:2010} and chiral optical selection rules influenced by the band spin splitting from strong spin-orbit coupling 
(SOC)~\cite{Xiao2012:PRL,Wang2012:NN,Chernikov2014:PRL,Chen2023:NC,Xu2014:NP}.
$K$ and $K'$ valleys, connected through time-reversal symmetry, are coupled to light with positive and negative helicity ($\sigma^+$, 
$\sigma^-$), where the SOC leads to the spin-valley locking. Excitons in TMDs provide a sensitive probe for 
proximity effects, from modified SOC and magnetism to charge-density wave and changing 
topology~\cite{Zutic2019:MT,Scharf2017:PRL,Thodika2025:X,Serati2023:NL,Cao2024:PRB,Cao2025:PRB,Faria2023:NM,Wu2017:PRL,Zhou2023:NM,Guo2024:PRB,Ruan2023:PNAS}. 
 
With a huge interest in altermagnets (AM)~\cite{Smejkal2022:PRX,Mazin2022:PRX,Bai2022:PRL,Krempasky2024:N,Song2025:NRM,Belashchenko2025:PRL,Denisov2024:PRBL,Khan2025:NPJ,Amin2024:N,Fernandes2024:PRB,Turan2025:X,Zhou2025:N,Zhu2025:PRL,Sun2024:PRB,Lin2025:PRL,Chen2025:PRL}, sharing vanishing magnetization with TMDs, but having broken time-reversal symmetry and spin splitting unrelated to SOC~\cite{Hayami2019:JPSJ,Hayami2020:PRB,Mazin2021:PNAS,Yuan2020:PRB,Yuan2021:PRM}, it would be important to elucidate the character of their excitons. Unlike many other magnets, AM offer a remarkable tunability of the spin splitting with alternating sign across the Brillouin zone~\cite{Duan2025:PRL,Gu2025:PRL,Urru2025:PRB,Zhu2025:SCP}. Several questions then naturally arise: (i) What are the underlying optical selection rules for excitons in AM? (ii) What is the implication of the versatile tunability of AM on their excitons and optical response? (iii) Could the excitonic properties offer an alternative signature of AM and complement the prior studies?

\begin{figure}
    \centering
    \includegraphics[width=1\linewidth]{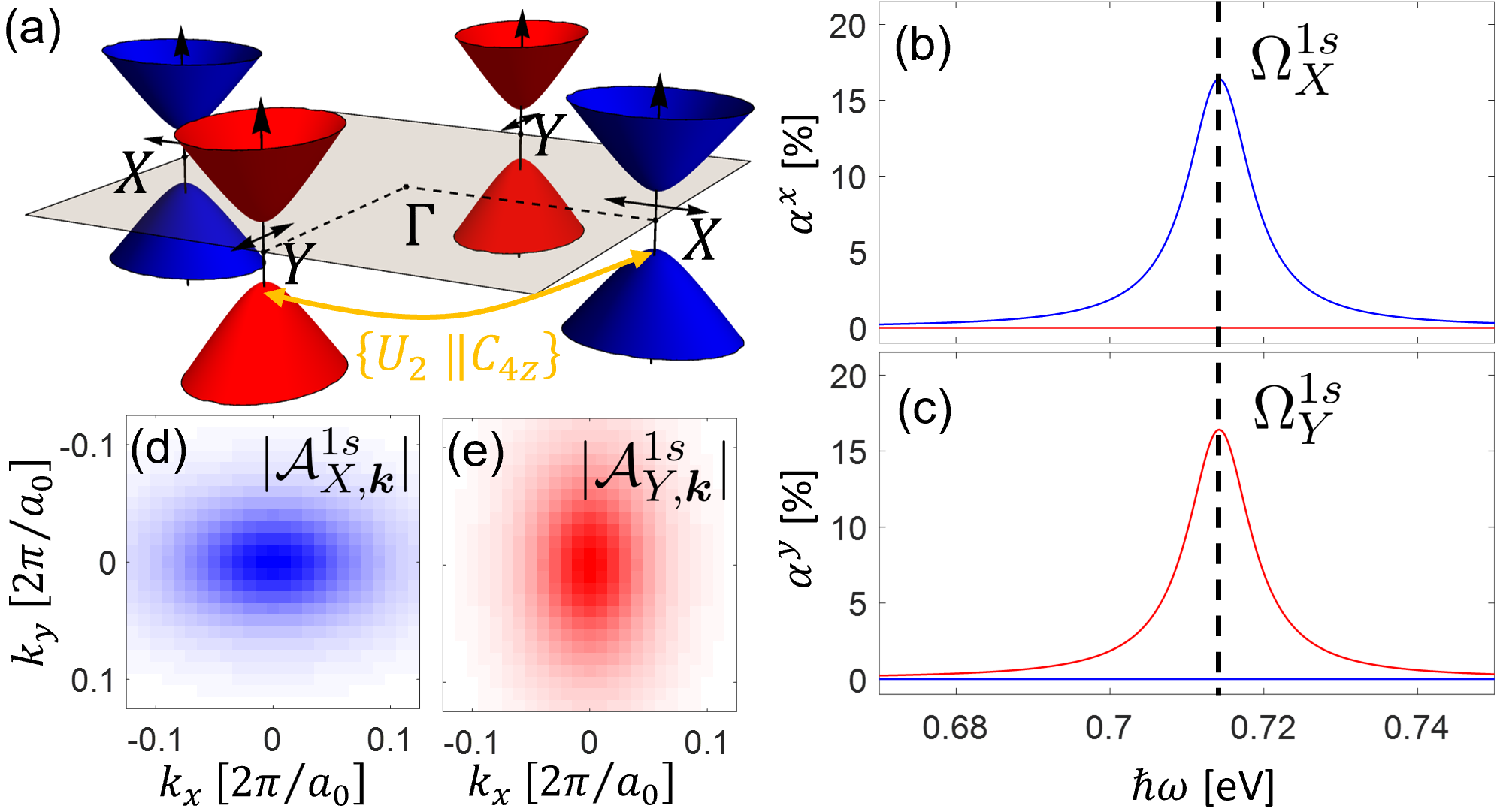}
    \caption{(a)~Schematic  
    of the band structure of a $d$-wave 2DAM with spin-polarized $X$ (blue) and $Y$ (red) valleys, connected by the symmetry operation $\{U_2||C_{4z}\}$. (b), (c)~Absorption spectra of $x$ and $y$ polarized light, with $\Omega_{X,Y}^{1s}$ the excitonic
    energies in the valleys $X$, $Y$,
    for the Case 1 (see text). (d), (e)~The corresponding momentum-space wave functions, 
   $\mathcal{A}^{1s}_{X,Y,\bm{k}}$, 
    of $1s$-like excitons in the $X$, $Y$ valleys, with $a_0$ the lattice constant and $(k_x,k_y)$ measured from $X$, $Y$.}
    \label{fig:XAM1}
\end{figure}

Building on the recent progress in 2DAM candidates~\cite{Ma2021:NC,Jiang2025:NP,Zhang2025:NP,Xu2025:X,Zhu2025:NL}, here we address these questions using a framework of the spin space group (SSG)~\cite{Brinkman1966:PRSA,Litvin1974:P,Liu2022:PRX,Chen2024:PRX,Xiao2024:PRX,Jiang2024:PRX,Yang2024:NC,Chen2025:N}, particularly suitable to study AM as well as a broader class of unconventional magnets~\cite{Shao2020:PRL,Liu2023:NL,Zhu2024:N,Liu2025:NP,Zhu2025:NC}. We focus on the $d$-wave 2DAM, shown in Fig.~\ref{fig:XAM1}(a), that host spin-polarized $X$ and $Y$ valleys. Unlike the $K$ and $K'$ valleys in TMDs, $X$ and $Y$ valleys are connected by the SSG symmetry operation $\{U_{2}||C_{4z}\}$,
which combines spin space $180^{\circ}$ rotation perpendicular to the collinear spin order and real space $90^{\circ}$ rotation along the z axis [Fig.~\ref{fig:XAM1} (a)]. Complementing SSG with the symmetry analysis of effective Hamiltonians and Bethe-Salpeter equation (BSE), we establish the excitonic optical selection rules for 2DAM, revealing alternating exciton wave function and $x/y$ linearly polarized light, that provide striking signatures for these magnets.
According to the SSG representations of the conduction and valence bands in the $X$ and $Y$ valleys, 2DAM can be classified into two cases (C1, C2), that separately hosts bright $s$-like  and $p$-like excitons. 

An exciton formed with electrons and holes from the valley $\tau$ is expressed as $|\tau,S\rangle=\sum_{\bm{k}}\mathcal{A}^{S}_{\tau,\bm{k}}\hat{c}^\dagger_{\tau, \bm{k}}\hat{c}^{\,}_{\tau, \bm{k}}|GS\rangle$, where $S$ labels the excitonic state, $\bm{k}$ is the wave vector measured from 
$\tau=X,Y$, $\mathcal{A}^S_{\tau,\bm{k}}$ is the $\bm{k}$-space exciton envelope wave function, $\hat{c}^\dagger_{\tau,\bm{k}}$~($\hat{c}_{\tau,\bm{k}}$) is the creation~(annihilation) operator of  the electron and $|GS\rangle$ is the ground state with fully occupied valence ($v$) 
and empty conduction ($c$) bands. The strength of an optical transition with polarization $\sigma$ to an exciton $|\tau,S\rangle$ is given 
by $\mathcal{M}^{\sigma}_{\tau, S}$
as
\begin{equation}\label{eq:select}
    \mathcal{M}^{\sigma}_{\tau, S}=\frac{1}{A}\sum_{\bm{k}}\mathcal{D}_{\tau,\bm{k}}^{\sigma}[\mathcal{A}^S_{\tau,\bm{k}}]^*,
\end{equation}
where $A$ is the 2D unit area, 
$\mathcal{D}_{\tau,\bm{k}}^{\sigma}=\langle \tau, c \bm{k}|\hat{v}_\sigma|\tau,v\bm{k}\rangle$, is the velocity matrix element, with $|\tau,c(v)\bm{k}\rangle$ for an electron~(hole) state.
The velocity operator, $\hat{v}_\sigma$, can be evaluated as 
$\hat{v}_{x/y} ={\partial {H}(\bm{k})}/{\partial( \hbar k_{x/y})}$, where ${H}(\bm{k})$ is the $\bm{k}$-dependent effective Hamiltonian, and $\hat{v}_{\pm}=(\hat{v}_x\pm i\hat{v}_y)/\sqrt{2}$. $\mathcal{A}^S_{\tau,\bm{k}}$ and the corresponding exciton energy $\Omega_{\tau}^S$ are obtained from
the BSE, see Supplemental Material~\cite{SM}, 
$\sum_{\bm{k}'}H_{\mathrm{BSE}}(\bm{k},\bm{k}')\mathcal{A}^S_{\tau,\bm{k}'}=\Omega_{\tau}^{S}\mathcal{A}_{\tau,\bm{k}}^S$, with the matrix element of the Hamiltonian~\cite{SM,Bechstedt:2016,Rohlfing2000:PRB,Scharf2016:PRB,Scharf2019:JPCM} defined as
\begin{equation}\label{eq:BSE}
    H_{\mathrm{BSE}}(\bm{k},\bm{k}')=E_{cv}(\bm{k})\delta_{\bm{k},\bm{k}'}-\frac{1}{A}V(|\bm{k}-\bm{k}'|)\langle c \bm{k}|c\bm{k}'\rangle\langle v \bm{k}'|v\bm{k}\rangle
\end{equation}
where $E_{cv}(\bm{k})=\epsilon_c(\bm{k})-\epsilon_v(\bm{k})$ is the $\bm{k}$-dependent energy difference between conduction and valence bands, the valley index is included in the band index $c(v)$, and $V(|\bm{q}|)$ is the Fourier component of the screened Coulomb interaction~\cite{Keldysh1979:JETP,Cudazzo2011:PRB}. 
While $\mathcal{A}_{\tau,\bm{k}}^S$ can be obtained by numerically solving the BSE, some important properties of excitons
can be obtained by the following symmetry analysis. 

In the 2D isotropic limit, $H_{\mathrm{BSE}}$ commutes with symmetry operation $C_{\infty z}$ around the $z$ axis, 
recovering a 2D hydrogen model where 
excitonic states are labeled by the principal 
and orbital angular momentum number~($1s$-, $2s$-, $2p$- like states)~\cite{Yang1991:PRA,Chao1991:PRB}. In TMDs, excitons in $K/K'$ valley  
resemble the physics of the 2D hydrogen, while the three-fold crystal field, band topology, the unusual 2D screening, and broken mirror symmetry lead to weak excitonic energy splitting and mixing~\cite{Glazov2017:PRB,Gong2017:PRB,Xu2020:PRL,Kormanyos2013:PRB,Chernikov2014:PRL,Zhang2018:PRL,Cao2024:PRB}. 
In the $X$ and $Y$ valleys of the $d$-wave 2DAM, $E_{cv}(\bm{k})$ has a strong anisotropy, related by the
$\{U_2||C_{4z}\}$ operation. This valley-dependent anisotropy originates from the inequivalent magnetic sublattices and is 
a key distinction between AM and their antiferromagnetic counterparts.
Compared with the isotropic limit, $H_{\mathrm{BSE}}$ has a 
reduced symmetry due to the anisotropy in $E_{cv}(\bm{k})$, analogous to the effect of a crystal field on an impurity ion~\cite{Dresselhaus:2007}. Excitons in phosphorene show how the anisotropy strongly influences excitonic properties~\cite{Rodin2014:PRB,Tran2014:PRB}. 

Both $\mathcal{D}_{\tau,\bm{k}}^{\sigma}$ and $\mathcal{A}_{\tau,\bm{k}}^S$ transform as irreducible representations (IRs) of the little group $G_{\tau}$~(denoted by  $\Gamma_{\mathcal{D^\sigma}}$ and $\Gamma_{\mathcal{A}}$). From Eq.~(\ref{eq:select}), $\mathcal{M}^{S}_{\tau,\sigma}$ transforms as the product
\begin{equation}\label{eq:MIR}
\Gamma_{\mathcal{D}^\sigma}\otimes\Gamma_{\mathcal{A}}^{\dagger }=(\Gamma^{\dagger}_{c}\otimes \Gamma_{\hat{v}_\sigma}\otimes\Gamma_{v})\otimes\Gamma_{\mathcal{A}}^{\dagger},
\end{equation}
where $\Gamma_{c(v)}$ and $\Gamma_{\hat{v}_\sigma}$ are the IRs of conduction~(valence) band and of light with $\sigma$ polarization.
The excitonic
optical selection rules 
can be read from Eq.~(\ref{eq:MIR}) as: An exciton is optically active under light with polarization $\sigma$, if $\Gamma_{\mathcal{D}^\sigma}\otimes\Gamma_{\mathcal{A}}^{\dagger}$ transform as the identity IR of the little group $G_{\tau}$ in valley $\tau$.
This establishes the excitonic optical selection 
rules
through group representation theory.

The symmetry of AM is described by the SSG, which allows independent spin 
and real space operations, in the limit of negligible SOC~\cite{Chen2024:PRX}. In our case, two spin-polarized $X$ and $Y$ valleys, as well as $\mathcal{M}_{X,S}^{\sigma}$ and $\mathcal{M}_{Y,S'}^{\sigma'}$, are connected by the SSG operation $\{U_2||C_{4z}\}$.

This $\{U_2||C_{4z}\}$ also connects $\sigma$ and $\sigma'$ in X and Y valleys, providing opportunities of coupled light polarization and valleys. $x/y$ light polarization can be coupled by $C_{4z}$, allowing possible valley polarization induced by $x/y$ polarized light, while $\sigma_\pm$ left/right circular polarization are invariant under $C_{4z}$.
Therefore, we focus on the optical response of $x/y$ polarized light $\mathcal{M}_{\tau,S}^{x/y}$ to take advantage of the valley degree of freedom. 
The relations $\mathcal{D}^{x/y}_{X,\bm{k}}=\mathcal{D}^{y/x}_{Y,\bm{k}}$ and $\mathcal{M}^{x/y}_{X,S}=\mathcal{M}^{y/x}_{Y,S'}$, with $|X,S\rangle$ and $|Y,S'\rangle$ form a pair of valley degenerate excitonic states, and their wave functions are protected by $\{U_2||C_{4z}\}$. 
                                      
Since the little group $G_{\tau}$ can be the direct product of its real space and spin parts, the IRs of $G_{\tau}$ also form the combinations of these two parts: 
$\Gamma^{\mathbb{R}}_i \Gamma^{\mathbb{S}}_j$~\cite{SM},
where $\Gamma^{\mathbb{R}}_i$ denotes the $i$-th IR of the space group formed by the pure real space operations of $G_{\tau}$; $\Gamma^{\mathbb{S}}_j$ denotes the IR of its spin-only group, with $j=\uparrow/\downarrow$ denotes the spin $\pm 1/2$ IR and $j=1$ denotes the identity IR. By separating the real and the spin space parts,
$\mathcal{M}^{S}_{\tau,\sigma}$ transforms as
$\Gamma^{\mathbb{R}}_{\mathcal{D}^\sigma}\Gamma^{\mathbb{S}}_{\mathcal{D}^\sigma}\otimes
{\Gamma^{\mathbb{R}}_{\mathcal{A}}}^\dagger{\Gamma^{\mathbb{S}}_{\mathcal{A}}}^\dagger = \allowbreak (\Gamma^{\mathbb{R}}_{\mathcal{D}^\sigma}\otimes{\Gamma^{\mathbb{R}}_{\mathcal{A}}}^\dagger)(\Gamma^{\mathbb{S}}_{\mathcal{D}^\sigma}\otimes{\Gamma^{\mathbb{S}}_{\mathcal{A}}}^\dagger)$.  The excitonic optical selection rules 
are now simplified to require $\mathcal{D}_{\tau,\bm{k}}^{\sigma}$ and $[\mathcal{A}^S_{\tau,\bm{k}}]^*$ transform as the same IR for both real 
and spin space parts.

We start from the spin
space part.
The selection rules 
require $\Gamma^{\mathbb{S}}_{\mathcal{D}^\sigma}\otimes{\Gamma^{\mathbb{S}}_{\mathcal{A}}}^{\dagger}=({\Gamma^{\mathbb{S}}_{c}}^\dagger\otimes \Gamma^{\mathbb{S}}_{\hat{v}_\sigma}\otimes\Gamma^{\mathbb{S}}_{v})\otimes{\Gamma^{\mathbb{S}}_{\mathcal{A}}}^{\dagger}=\Gamma^{\mathbb{S}}_1$. Since both $\Gamma^{\mathbb{S}}_{\hat{v}_\sigma}$ and $\Gamma^{\mathbb{S}}_{\mathcal{A}}$ are identity IR of the spin-only group, the selection rules 
are
simplified to ${\Gamma^{\mathbb{S}}_c}^{\dagger}\otimes\Gamma^{\mathbb{S}}_{v}=\Gamma^{\mathbb{S}}_1$, which is equivariant to $\Gamma^{\mathbb{S}}_c=\Gamma^{\mathbb{S}}_v=\Gamma^{\mathbb{S}}_{\uparrow/\downarrow}$. 
In summary, we recover the spin selection rules
that require 
bright excitons to be formed with electrons and holes with the same spin.

\begin{table}
\caption{\label{tab:IR} Multiplication relations and basis functions of $\Gamma^{\mathbb{R}}_i$ of the little group $^12^1m^1m^{\infty m}1$.}
\begin{ruledtabular}
    \begin{tabular}{c|cccc|c}
&$\Gamma^{\mathbb{R}}_1$&$\Gamma^{\mathbb{R}}_2$&$\Gamma^{\mathbb{R}}_3$&$\Gamma^{\mathbb{R}}_4$&Functions\\ \hline $\Gamma^{\mathbb{R}}_1$&$\Gamma^{\mathbb{R}}_1$&$\Gamma^{\mathbb{R}}_2$&$\Gamma^{\mathbb{R}}_3$&$\Gamma^{\mathbb{R}}_4$&$z,x^2+y^2,x^2-y^2,z^2$\\
$\Gamma^{\mathbb{R}}_2$&&$\Gamma^{\mathbb{R}}_1$&$\Gamma^{\mathbb{R}}_4$&$\Gamma^{\mathbb{R}}_3$&$R_z,xy$\\
 $\Gamma^{\mathbb{R}}_3$&&&$\Gamma^{\mathbb{R}}_1$&$\Gamma^{\mathbb{R}}_2$&$x,xz$\\
 $\Gamma^{\mathbb{R}}_4$&&&&$\Gamma^{\mathbb{R}}_1$&$y,yz$\\
\end{tabular}
\end{ruledtabular}
\end{table}

\begin{table*}
\caption{\label{tab:s-rule}The excitonic optical selection rules,
with $\checkmark$ for non-zero quantity. 
For the same type of Hamiltonian but different valleys, exciton energies are degenerate and non-zero excitonic optical transitions are equal~[in the same cell of the table]. For example, $\Omega^{2p_y}_X=\Omega^{2p_x}_Y$ and $\mathcal{M}^x_{X,2p_y}=\mathcal{M}_{Y,2p_x}^y$. }
\begin{ruledtabular}
    \begin{tabular}{cc||cc|cc|cc}
&$\sigma$&$\mathcal{M}_{X,1s}^{\sigma}$&$\mathcal{M}_{Y,1s}^{\sigma}$&$\mathcal{M}_{X,2p_y}^{\sigma}$&$\mathcal{M}_{Y,2p_x}^{\sigma}$&$\mathcal{M}_{X,2p_x}^{\sigma}$&$\mathcal{M}_{Y,2p_y}^{\sigma}$\\ \hline 
  Case 1 [Eq.~(\ref{eq:s-H})]     
  &$x$ &$\checkmark$&$0$&0&0&0&0\\
 &$y$&0&$\checkmark$&0&0&0&0\\
\hline
  Case 2 [Eq.~(\ref{eq:p-H})]
  &$x$ &0&0&$\checkmark$&0&0&$\checkmark$\\
 &$y$ &0&0&0&$\checkmark$&$\checkmark$&0\\
\end{tabular}
\end{ruledtabular}
\end{table*}

Next, we examine
the real space part. This part varies in different $d$-wave 2DAM, allowing more complex and interesting selection rules, as compared to the spin space part. As an example, we choose $G_{\tau}=\,^12^1m^1m^{\infty m}1$, where the real space part is the point group $2mm$. The multiplication relations and basis functions of the real space IR $\Gamma^{\mathbb{R}}_i$ are summarized in Table~\ref{tab:IR}. 
We 
see from the table that $s$-, $p_x$-, $p_y$- and $d_{xy}$- excitons transform as $\Gamma^{\mathbb{R}}_1$, $\Gamma^{\mathbb{R}}_3$, $\Gamma^{\mathbb{R}}_4$, and $\Gamma^{\mathbb{R}}_2$, while $\hat{v}_{x/y}$ transforms as $\Gamma^{\mathbb{R}}_3 /\Gamma^{\mathbb{R}}_4$.
In the following, we study excitonic optical selection rules for Case 1 and Case 2 (C1 and C2) of the combination of orbital of conduction and valence bands characterized by $\Gamma_{c}^{\dagger}\otimes\Gamma_v$.

In C1, the real space part of $\Gamma_{c}^{\dagger}\otimes\Gamma_v$ is $\Gamma^{\mathbb{R}}_3$ or $\Gamma^{\mathbb{R}}_4$. As an example, 
we examine 
where $\Gamma_{c}^{\dagger}\otimes\Gamma_v$ transforms as $\Gamma^{\mathbb{R}}_3\Gamma^{\mathbb{S}}_{\uparrow}$ in the $X$ valley, $\Gamma^{\mathbb{R}}_4\Gamma^{\mathbb{S}}_{\downarrow}$ in the $Y$ valley.
$\mathcal{D}^{x/y}_{X,\bm{k}}$ transforms as $\Gamma^{\mathbb{R}}_1\Gamma^{\mathbb{S}}_\uparrow$/$\Gamma^{\mathbb{R}}_2\Gamma^{\mathbb{S}}_\uparrow$, while $\mathcal{D}^{x/y}_{Y,\bm{k}}$ transforms as $\Gamma^{\mathbb{R}}_2\Gamma^{\mathbb{S}}_\downarrow$/$\Gamma^{\mathbb{R}}_1\Gamma^{\mathbb{S}}_\downarrow$.
We can see that both $s$- and $d_{xy}$-like excitons should be bright as they transform as $\Gamma^{\mathbb{R}}_1$ and $\Gamma^{\mathbb{R}}_2$, but the brightness of a $d$-like exciton is much weaker than an $s$-like exciton, due to a much smaller wave amplitude. Therefore, the excitonic optical selection rules are 
simplified as $s$-like excitons in $X$/$Y$ valley are coupled to $x$/$y$ polarized light. This rule is reflected in Table~\ref{tab:s-rule}.  
One can verify that by exchanging the orbital characters of the $X$ and $Y$ valleys, the polarization and valley index coupling are exchanged.  

In C2, the real space part of  $\Gamma_{c}^{\dagger}\otimes\Gamma_v$ transforms as $\Gamma^R_1$ or $\Gamma^R_2$. As an example, we present the case where  
$\Gamma_{c}^{\dagger}\otimes\Gamma_v$ is $\Gamma^{\mathbb{R}}_2$ in both valleys.
$\mathcal{D}^{x/y}_{\tau, \bm{k}}$ transforms as $\Gamma^{\mathbb{R}}_4$/$\Gamma^{\mathbb{R}}_3$ in both valleys. 
According to the excitonic optical selection rules, $2p_y$- and $2p_x$-like excitons are bright under $x$ ($y$) 
polarized light, since their wave functions transform as $\Gamma^{\mathbb{R}}_4$  
($\Gamma^{\mathbb{R}}_3$). 
Together with the $\{U_2||C_{4z}\}$ transformation that connects the exciton wave functions of a 
valley-degenerate exciton pair,
$\mathcal{A}^{2p_{x/y}}_{X,\bm{k}}$ and $\mathcal{A}^{2p_{y/x}}_{Y,\bm{k}}$, we get an alternating optical selection rules 
of excitons. 
One example is that $2p_y$-like exciton in the $X$ valley and the $2p_x$-like exciton in $Y$ valley are optically active to $x$ and $y$ polarized light separately, with the same excitonic energy $\Omega^{2p_y}_X=\Omega^{2p_x}_Y$ and optical absorption strength 
$\mathcal{M}^{x/y}_{X,2p_y}=\mathcal{M}^{y/x}_{Y,2p_x}$. 
A summary is shown in Table~\ref{tab:s-rule}. 

So far, we have established the excitonic optical selection through symmetry analysis. 
We verify these predictions and get more insights by 
solving the BSE  
to obtain
the measurable
absorption of linearly-polarized light as a function of its frequency, $\omega$
\begin{equation}\label{eq:abs}
    \alpha^{x/y}(\omega)=\frac{4e^2\pi^2}{c \,\omega}\sum_{S,\tau=X,Y}|\mathcal{M}^{x/y}_{\tau,S}|^2\delta(\hbar\omega-\Omega_{\tau}^{S}),
\end{equation}
where $c$ is 
speed of light, $\delta$ is the delta function, represented by the 5$\;$meV broadening~\cite{SM}. 
The corresponding single-particle energies and states are 
obtained 
from low-energy effective Hamiltonians, which are constructed through $\bm{k}\cdot\bm{p}$ method considering symmetry constrains and corresponding $\Gamma_{c,v}$ in C1, C2~\cite{Liu2024:npjQM}.
The selection of parameters is 
given in~\cite{SM}, to evaluate 
$\mathcal{M}^{x/y}_{\tau,S}$. 

For C1, the low-energy effective Hamiltonian is 
\begin{equation}\label{eq:s-H}
\begin{split}
        H^{\mathrm{C1}}_{X(\uparrow)}&=(\Delta+\beta_1k_x^2+\beta_2k_y^2)\sigma_z+\gamma k_x \sigma_y,\\
        H^{\mathrm{C1}}_{Y(\downarrow)}&=(\Delta+\beta_1k_y^2+\beta_2k_x^2)\sigma_z+\gamma k_y \sigma_y,
\end{split}
\end{equation}
where, $k_{x,y}$ are the components of wave vectors measured from the band extremes at the $X$ and $Y$ points, $\bm{\sigma}$ denotes the Pauli matrices that act on a basis of conduction and valence band edge states. 
The difference between $\beta_1$ and $\beta_2$ represents the anisotropy of the conduction and valence bands, originating from the anisotropy of 
the crystal field on magnetic sublattices. $\gamma$ originates from the hybridization of the conduction and valence bands.
An alternative realization of the C1 Hamiltonian is realized by the exchange $\gamma k_x\sigma_y \leftrightarrow\gamma k_y\sigma_y$.
The single-particle energy and state is obtained by solving 
$H^{\mathrm{C1}}_{\tau(\uparrow/\downarrow)}|\tau,n\bm{k}\rangle=\epsilon_n|\tau,n\bm{k}\rangle$ with $n=v,c$ for the valence, conduction band. The velocity matrix elements can be obtained directly as $\mathcal{D}^x_{X,\bm{k}}=\mathcal{D}^y_{Y,\bm{k}}=\gamma$ and $\mathcal{D}^y_{X,\bm{k}}=\mathcal{D}^x_{Y,\bm{k}}=0$.
By solving the BSE in Eq.~(\ref{eq:BSE}),
we get the absorption spectrum as in Figs.~\ref{fig:XAM1}(b), (c). As shown in Figs.~\ref{fig:XAM1}(d), (e), the corresponding exciton wave functions are $s$-like with alternating anisotropy in the two valleys. 
Our numerical results confirm that $1s$-exciton in the $X$/$Y$ valley is coupled to $x$/$y$ polarized light as obtained from the group theory analysis. 
We also confirm that exciton wave functions 
$\mathcal{A}^{1s}_X$  and $\mathcal{A}^{1s}_Y$ are connected 
by $\{U_2||C_{4z}\}$, as predicted from our symmetry analysis.

\begin{figure}
    \centering
    \includegraphics[width=0.95\linewidth]{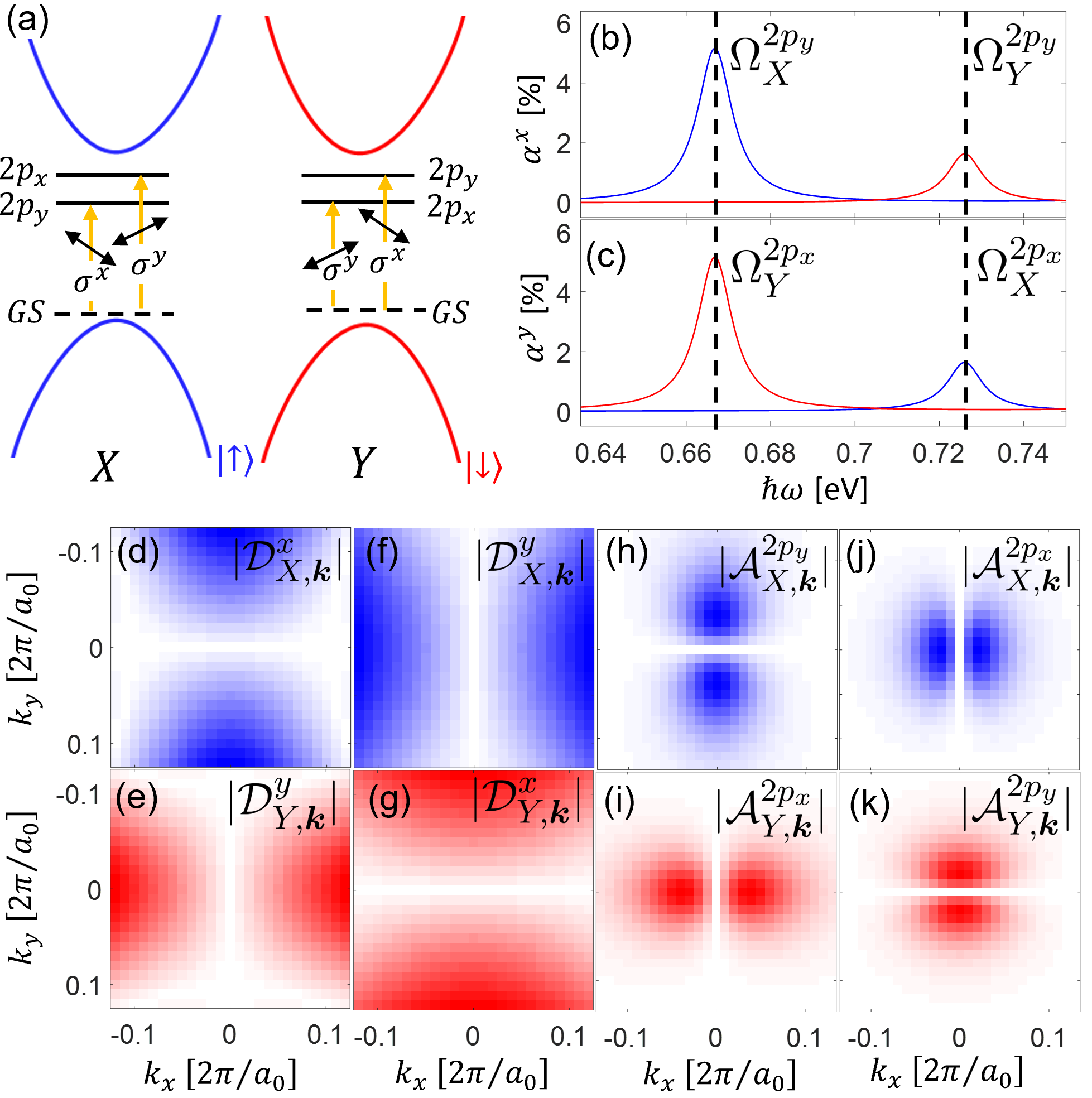}
    \caption{Case 2, with the $X$ (blue) 
    and $Y$ (red) valleys. (a)~Schematic 
    of the excitonic optical selection rules. (b), (c)~Absorption spectra of $x$ and $y$ polarized light,
    with a single-particle
    band gap $2\Delta=0.82\;$eV.
    (d)-(g)~The velocity matrix elements $|\mathcal{D}^{x/y}_{\tau,\bm{k}}|$. (h)-(k)
    The exciton wave function, with $2p_{y/x}$-like exciton in the $X/Y$ valley and $2p_{y/x}$-like exciton in the $Y/X$ valley corresponds to the $x/y$ absorption of the first and second peak as in (b), (c). In (d)-(g), ($k_x$, $k_y$) is measured from the corresponding $X$ or $Y$ valley.} 
    \label{fig:XAM2}
\end{figure}

For C2, the low-energy effective Hamiltonian is
\begin{equation}\label{eq:p-H}
\begin{split}
        H^{\mathrm{C2}}_{X(\uparrow)}&=(\Delta+\beta_1 k_x^2+\beta_2 k_y^2)\sigma_z+\gamma' k_x k_y\sigma_x,\\
        H^{\mathrm{C2}}_{Y(\downarrow)}&=(\Delta+\beta_1 k_y^2+\beta_2 k_x^2)\sigma_z+\gamma' k_x k_y\sigma_x.
\end{split}
\end{equation}
Compared to C1, the changes in the IRs of the conduction and valence bands lead to different forms of hybridization term $\gamma'$.
An alternative C2 Hamiltonian is realized by replacing the second term by $(\gamma_1 k_x^2+\gamma_2 k_y^2)\sigma_x$ or $(\gamma_1 k_y^2+\gamma_2 k_x^2)\sigma_x$ for the $X$ or $Y$ valley Hamiltonian.
The velocity matrix elements in the $X$ and $Y$ valleys are connected by $\{U_2||C_{4z}\}$, to satisfy 
$\mathcal{D}^x_{X,\bm{k}}=-\mathcal{D}^y_{Y,\bm{k}}=\gamma' k_y$ and 
$\mathcal{D}^y_{X,\bm{k}}=-\mathcal{D}^x_{Y,\bm{k}}=\gamma' k_x$ agreeing with 
our calculations 
in Figs.~\ref{fig:XAM2}(d)-(g).
In Figs.~\ref{fig:XAM2}(b), (c), 
there are
two peaks in the absorption spectra of $x$- and $y$-polarized light,
with alternating coupling to $X$ and $Y$ valleys. Symmetries of excitons are identified by plotting their wave functions as shown in Figs.~\ref{fig:XAM2}(h)-(k). 
Therefore, we find that the excitonic states $|X(Y), 2p_{y(x)}\rangle$ and $|Y(X), 2p_{y(x)}\rangle$ are responsible for the first and second peaks in the absorption 
of $x$ ($y$) polarized light, respectively.
The excitonic binding energies of 100 meV or more, which can be also inferred from the 
single-particle gap of 0.82$\;$eV, 
are encouraging for the
room-temperature robustness of excitons.
The alternating excitonic optical selection rules are 
summarized in Table.~\ref{tab:s-rule}.
In Figs.~\ref{fig:XAM2}(b), (c), 
the first peaks are higher than the second, indicating the relationship 
$\mathcal{M}^x_{X,2p_y}>\mathcal{M}^y_{X,2p_x}$. This is the result of the anisotropy of the exciton 
wave functions; accordingly,  $\mathcal{A}^{2p_{y}}_{X,\bm{k}}$ and $\mathcal{A}^{2p_{x}}_{X,\bm{k}}$
are stretched along their $p$-wave lobe direction and vertical direction, respectively~[Figs.~\ref{fig:XAM2}(h)-(k)], due to the anisotropic $E_{vc}(\bm{k})$~\cite{Rodin2014:PRB,Tran2014:PRB}. 
As a result, the former is larger when multiplied by $\mathcal{D}^{x/y}_{X,\bm{k}}$, 
which disperses linearly in $k$. 
We expect a reduced/enhanced anisotropy of the band dispersion by adjusting $\beta_x/\beta_y$
to make the heights of two peaks closer/further. This can be achieved by strain or electric field. The excitonic optical selection rules of both types are summarized in Table.~\ref{tab:s-rule}.

\begin{figure}
    \centering
    \includegraphics[width=0.95\linewidth]{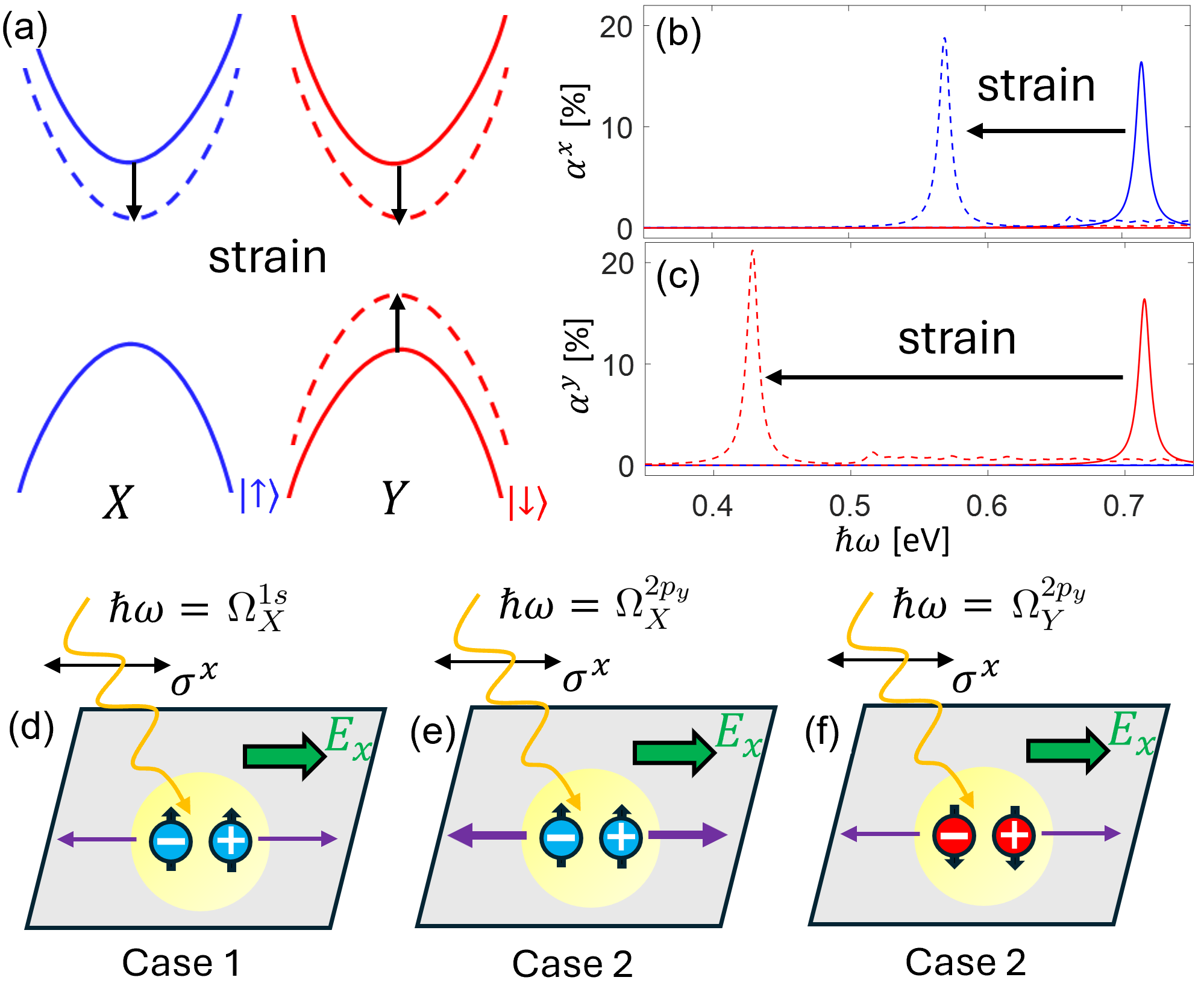}
    \caption{(a)~Schematic of the band edge motion under strain. (b), (c)~Evolution of the absorption spectra under strain. Splitting of the peaks with different polarization reflects the broken valley degeneracy. (d)-(f)~Anisotropic valley-polarized photocurrent with optical excitation and electric field.}
    \label{fig:XAM3}
\end{figure}

We next discuss how to lift the valley degeneracy of excitonic energies. 
In TMDs, an application of the magnetic field leads only to a small Zeeman splitting of 
$\sim$0.1-0.2~meV/T~\cite{Zutic2019:MT}, which can be enhanced up to $16$~meV/T by proximity effect from magnetic 
substrates~\cite{Zutic2019:MT,Norden2019:NC}. 
Instead, another mechanism is particularly important for 2DAM: Strong anisotropy of the magnetic sublattices leads to a significant valley splitting upon the application of strain. An efficient valley splitting of $0.15$~eV under $5\%$ uniaxial strain for $\mathrm{V}_2\mathrm{Se}_2\mathrm{O}$ is obtained,  as illustrated in Fig.~\ref{fig:XAM3}(a)~\cite{Ma2021:NC}. 
By calculating the evolution of absorption spectra under strain based on C1 Hamiltonian in Eq.~(\ref{eq:s-H}), 
we show that
a redshift of $y$-polarized absorption is 0.15~eV larger than the energy shift for $x$ polarized light, 
in Figs.~\ref{fig:XAM3}(b), (c). 
This shift reflects the strain-induced difference of the exciton energies in the $X$ and $Y$ valleys.
The obtained energy shifts originate mostly from the strain-induced change in the single-particle band edge energy: $\Delta\Omega^S=\Delta E_{cv}(\bm{k}=0)$, with negligible role of strain-dependent changes in the exciton binding energy.
Similar effects can also be achieved by inducing valley splitting through the proximity 
effect~\cite{Zutic2019:MT,Zhu2025:X,Zhu2025:X-SFETs} or external fields~\cite{Vila2025:PRB}.
The exciton-dominated absorption spectra evolution thus provides a fingerprint of the efficient lifting of the valley degeneracy, and guidance to strain engineering of optoelectronic devices.

The spin-valley locking in the $d$-wave 2DAM along with the valley-sensitive selection rules for the optical absorption imply the realization of interesting valley-spin-polarized anisotropic photocurrents, 
illustrated in Figs.~\ref{fig:XAM3}(d)-(f).
With a photoexcitation of an appropriate frequency applied to the AM sample generates excitons which then disassociate to electrons and holes by the applied in-plane electric field with the formation of the electrical current. Let us consider the photocurrents for an applied electric field, $E_x$, along $x$.
In C1, the light with angular frequency $\Omega_{\tau}^{1s}/\hbar$ and linear polarization along $x$ generates excitons only in the $X$ valley, with their subsequent conversion to the photocurrent $j_x = \sigma_{xx}^{\tau} E_x$, as shown in Fig.~\ref{fig:XAM3}(d), where $\sigma_{xx}^{X}$ is the photoconductivity in $X$ valley along the $x$ direction.
 Choosing the linear polarization along $y$ will produce the photocurrent in the $Y$ valley, now with a different magnitude of the current $j_x = \sigma_{xx}^Y E_x$, due to anisotropic band structure of AM ($\sigma_{xx}^X \neq \sigma_{xx}^Y$).
In C2 AM, 
we can realize a similar situation without changing the linear polarization of the incident light, but by tuning the resonant excitation frequency instead.
For the $x$ polarization of light,
we get the photocurrent, $j_x^X$,
in the $X$ valley for $\Omega^{2p_y}_X$ while taking different $\Omega^{2p_y}_Y$ will give us the photocurrent, $j_x^Y$, in the $Y$ valley, as shown in Figs.~\ref{fig:XAM3}(e)-(f).
These photocurrents have different magnitudes, $j_x^X \neq j_x^Y$, due to the band structure anisotropy. 
The same notation is used in Figs.~\ref{fig:XAM1}(b), (c) and Figs.~\ref{fig:XAM2}(b), (c).

Following the theoretical prediction of 
$\mathrm{V_2Te_2O}$ and $\mathrm{V_2Se_2O}$, there is rapid progress in the first-principles calculations and experiments that expand, study, and experimentally confirm the family of $d$-wave 2DAM with spin-polarized $X/Y$ valleys, providing opportunities to test our 
predictions of exciton properties in real 
materials~\cite{Ma2021:NC,Jiang2025:NP,Zhang2025:NP,Xu2025:X,Zhu2025:NL}.
However, due to the additional constraint imposed by 
the space inversion symmetry on exciton selection rules, such systems are often unable to host bright excitons~\cite{SM}.
Breaking the out-of-plane mirror symmetry of $\mathrm{V_2Se_2O}$, e.g., through Janus monolayers, substrates, or electric field, can modify the valley properties consisting of our C1 (Eq.~(\ref{eq:s-H}) and \cite{SM}). Furthermore, to identify additional materials with desirable excitonic properties, we screened compounds sharing the same structures as the $\mathrm{V_2Se_2O}$ family and their Janus derivatives. Materials exhibiting energy valleys at the $X$ and $Y$ points are summarized in Tables S2 and S3~\cite{SM}, five each for C1 and C2.
These material candidates provide promising platforms to 
experimentally verify our predictions.
Our work also provides a guidance for future material screening with alternating excitonic optical selection rules.

While SOC is not included in our theory, the large valley spin splitting from AM order makes spins robust against SOC, and justify its neglect. 
Even though our study focuses on the $X/Y$ valley physics of $d$-wave 2DAM, this generic framework that relies on the fundamental group theory can also be generalized to higher-order wave AM. However, these systems lack spin-polarized high-symmetry valleys and do not possess $C_4$ rotation to link the two spin channels and ensure full spin-valley-polarization locking (see 
~\cite{SM}), thus they are less suitable for our exciton framework.

Future studies of 2DAM may take advantage of this light-valley coupling as the tunability by uniaxial strain was also employed for ultrafast polarization switching 
and information transfer in spin-lasers~\cite{Lindemann2019:N}. With their complex
spin textures~\cite{Hu2025:PRX}, these 2DAM may also advance opportunities for an unconventional
spin-orbit torque to integrate room-temperature spintronics, electronics, and photonics at zero applied magnetic field~\cite{Dainone2024:N}.  Considering strong correlation effects, AM can also serve as a promising platform for realizing an excitonic insulator with alternating spin polarization~\cite{Guo2023:arXiv,Sheng2025:arXiv}
.

In summary, our work develops a general symmetry-based framework for classifying and predicting excitonic optical selection rules in 2DAM. We uncover alternating bright exciton states exhibiting spin, valley, and polarization locking, along with experimentally accessible optical signatures, supported by BSE calculations and effective Hamiltonian. Materials candidates predicted by our symmetry framework, together with first-principles calculations, offer promising platforms for realizing excitons in 2DAM. Due to the alternating distribution of the exciton states in two valleys, these properties provide fingerprints of AM and complement prior detection approaches mainly relying on transport measurements. Extensions of our studies to highly-tunable and inherently anisotropic AM could enable using excitons and anisotropic photocurrent to probe unexplored topological phases~\cite{Zhou2023:NM} and changes in the band topology~\cite{Xu2020:PRL}. Our approach is applicable to other AM, not just 2DAM, and unconventional magnets, further enriched by the prospect of altermagnetic proximity effects~\cite{Zhu2025:X} as well coupling of excitons with other collective excitations~\cite{Grzeszczyk2023:AM,VanTaun2017:PRX,Shao2025:NM}.

\textit{Note added}---Following the completion of this work, we became aware of excitonic studies in AM using first-principles calculations~\cite{Wang2025:arXiv,Sun2025:arXiv}.

This work is supported by the U.S. Department
of Energy, Office of Science, Basic Energy Sciences under Award No. DE-SC0004890. Computational resources were provided by the UB Center for Computational Research. 

\bibliography{xam}

\end{document}